\documentstyle[psfig]{mn}%

\def\atp{a_{\rm tp}}
\def\ap{a_{\rm p}}
\def\ntp{n_{\rm tp}}
\def\np{n_{\rm p}}
\def\einst{e_{\rm inst}}
\def\emax{e_{\rm max}}
\def\iinst{i_{\rm inst}}
\def\imax{i_{\rm max}}
\def\ds{\displaystyle}

\def\i{\relax\ifmmode{\rm i}\else\char16\fi}

\def\lesssim{{_ <\atop{^\sim}}}
\def\lta{\lesssim}

\def\lesssim{\mathrel{\hbox{\rlap{\hbox{\lower4pt\hbox{$\sim$}}}\hbox{$<$}}}}
\def\gtrsim{\mathrel{\hbox{\rlap{\hbox{\lower4pt\hbox{$\sim$}}}\hbox{$>$}}}}

\begin{document}

   \title[Structure of Possible Long-lived Asteroid Belts]
	{Structure of Possible Long-lived Asteroid Belts}

   \author[Evans N.W. \& Tabachnik S.A.]
          {N.W. Evans$^1$ and S.A. Tabachnik$^2$,\\
           $^1$ Theoretical Physics, 1 Keble Road, Oxford OX1 3NP \\
	   $^2$ Princeton University Observatory, Peyton Hall,
           Princeton, NJ 08544-1001, USA
           }
\date{Received ...; accepted ...}

\maketitle

\begin{abstract}
High resolution simulations are used to map out the detailed structure
of two long-lived stable belts of asteroid orbits in the inner Solar
system. The Vulcanoid belt extends from 0.09 to 0.20 astronomical
units (au), though with a gaps at 0.15 and 0.18 au corresponding to
de-stabilising mean motion resonances with Mercury and Venus.  As
collisional evolution proceeds slower at larger heliocentric
distances, kilometre-sized or larger Vulcanoids are most likely to be
found in the region between $0.16$ and $0.18$ au.  The optimum
location in which to search for Vulcanoids is at geocentric ecliptic
longitudes $9^\circ \le |\ell_{\rm g} | \le 10^\circ$ and latitudes $|
\beta_{\rm g} | < 1^\circ$.  Dynamically speaking, the Earth-Mars belt
between 1.08-1.28 au is an extremely stable repository for asteroids
on nearly circular orbits. It is interrupted at 1.21 au due to the
3:4 commensurability with the Earth, while secular resonances with
Saturn are troublesome beyond 1.17 au.  These detailed maps of the
fine structure of the belts can be used to plan search methodologies.
Strategies for detecting members of the belts are discussed, including
the use of infrared wide-field imaging with VISTA, and forthcoming
European Space Agency satellite missions like {\it GAIA} and {\it
BepiColombo}.
\end{abstract}


\section{Introduction} 

The most abundant minor bodies of the Solar system are the asteroids
whose Main Belt extends between 2 and 3.5 au. The asteroids are
remnant planetesimals. It has often been suggested that the Main Belt
is perhaps the only zone in the Solar System where planetesimals can
survive in long-lived orbits for ages of the order of 4.5 Gyrs (e.g.,
Chapman 1987; Laskar 1997).  Recently, however, Holman (1997)
uncovered evidence for a possible belt between Uranus and Neptune by
4.5 Gyr integrations of test particles in the gravity field of the Sun
and the four massive planets. Mikkola \& Innanen (1995) performed a 3
Myr test particle calculation in the inner Solar system which
suggested the existence of narrow bands of stability between each of
the terrestrial planets.  Subsequently, Evans \& Tabachnik (1999)
reported the results of 100 Myr test particle
calculations. Extrapolating the results to the age of the Solar
system, they conjectured that there might be two long-lived belts of
stable asteroids in the inner Solar system. These are the domain of
the Vulcanoids between 0.09 and 0.20 astronomical units (au) and the
Earth-Mars belt between 1.08 and 1.28 au respectively.

In this {\it Letter}, we report results on the structure of the belts
inferred from high resolution numerical simulations. The general
procedure is similar to many recent studies on the stability of test
particles in the Solar system.  At semimajor axes separated by 0.002
au in each of the two belts, five test particles are launched on
initially circular orbits in the ecliptic with starting longitudes $n
\times 72^\circ$ with $n = 0, \dots,4$.  The test particles are
perturbed by the Sun and planets, but do not themselves exert any
gravitational forces. The full effects of all the planets (except
Pluto) are included. The initial positions and velocities of the
planets, as well as their masses, come from the JPL Planetary and
Lunar Ephemerides DE405 and the starting epoch is JD 2440400.5 (28
June 1969). After each timestep, the test particles are examined. If
their orbits have become hyperbolic, or if they have entered the
sphere of influence of any planet or if they have approached closer
than ten solar radii to the Sun, then they are removed. The
simulations are run for 100 Myr. The calculations for the 300
Vulcanoid test particles are performed in long double precision to
minimise round-off error. The 500 test particles in the Earth-Mars
belt are simulated using double precision on the Oxford Supercomputer.
Even using a fast symplectic integrator with individual timesteps
(Saha \& Tremaine 1994), the calculations still consumed several
months.
  
\begin{figure*}
\centerline{\psfig{file=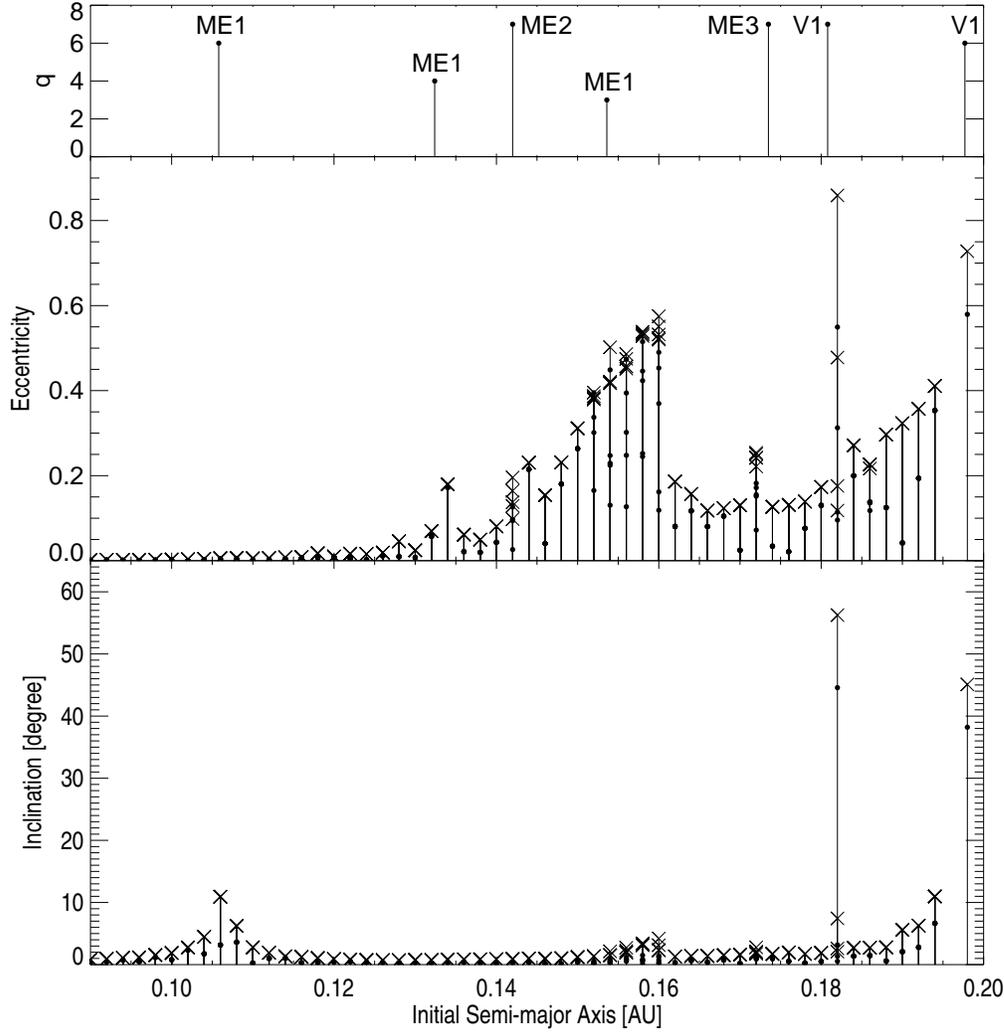,width=0.67\hsize}}
\caption{The two lower panels show the eccentricity and the
inclination of the remaining test particles in the Vulcanoid belt
after 100 Myr. Filled circles represent the instantaneous values,
while crosses indicate the maximum values over the integration
timespan. The location of the mean motion resonances are plotted in
the topmost panel according to the order of the commensurability
$q$. The notation ME and V refers to Mercury and Venus respectively,
while the number following it is the value of $k$ in
equation~(\ref{eq:res}).
\label{fig:figone}
}
\end{figure*}
\begin{figure*}
\centerline{\psfig{file=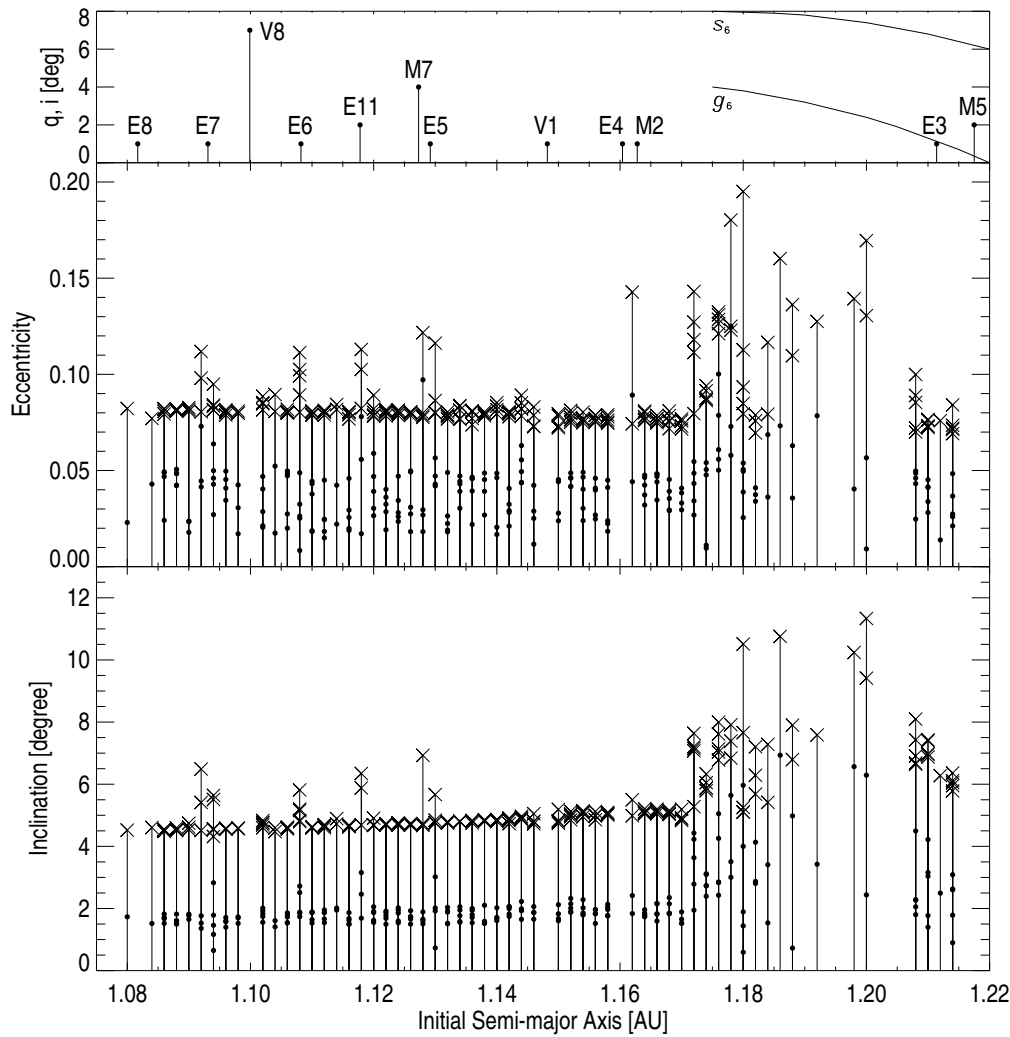,width=0.67\hsize}}
\caption{The two lower panels illustrate the eccentricity and the
inclination of the remaining test particles in the Earth-Mars belt
after 100 Myr. Again, filled circles and crosses represent the
instantaneous and the maximum values. The upper panel shows the
location of the mean motion and secular resonances. The notation V, E
and M refers to Venus, the Earth and Mars. From 1.175 au to 1.22 au,
the two arcs labelled $g_6$ and $s_6$ mark the locations at which two
linear secular resonances due to Saturn are prominent as a function of
semimajor axis and inclination. We note that the stable region between
1.208 and 1.214 au is interrupted at 1.212 au due to the 3:4
commensurability with the Earth.
\label{fig:figtwo}
}
\end{figure*}

\section{The Vulcanoids}

The Vulcanoids are a population of intra-Mercurial bodies originally
proposed by Weidenschilling (1978) to resolve seeming contradictions
in the geological and cratering history of Mercury.  The inner edge of
the Vulcanoid belt is $\sim 0.09$ au. Small bodies closer than this to
the Sun are removed as a consequence of both Poynting-Robertson drag
and the Yarkovsky effect (e.g., Vokroulick\'y, Farinella \& Bottke
2000). Evans \& Tabachnik (1999) argued that the outer edge is $0.20$
au. Small bodies closer than this to Mercury evolve swiftly into the
planet's sphere of influence.  The robustness of orbits in the
Vulcanoid belt stems from the fact that there is only one bounding
planet and so it is analogous to the stability of the Kuiper-Edgeworth
belt.  Even after 100 Myr, some $80 \%$ of the Vulcanoid orbits still
have eccentricities $e < 0.2$ and inclinations $i < 10^\circ$ in our
high resolution simulation. The two lower panels of
Figure~\ref{fig:figone} illustrate the eccentricities and inclinations
of the surviving test particles at the end of the integration. In the
interval $0.09 \le a_{\rm tp} \le 0.20$, the averaged values for the
eccentricity are $\langle \einst \rangle = 0.0935$ and $\langle \emax
\rangle = 0.15$ and for the inclination $\langle \iinst \rangle =
1.21^\circ$ and $\langle \imax \rangle = 5.36^\circ$, the subscripts
referring to the instantaneous and maximum quantities respectively.
The upper panel indicates the location of the mean motion resonances,
which occur when the orbital period of one planet $\np \sim
\ap^{-3/2}$ and a test particle $\ntp \sim \atp^{-3/2}$ are
commensurable, or close to the ratio of small integers
\begin{equation}
{\ntp \over \np} = \cases{ {\ds k+q \over \ds k} & if $\atp < \ap$,\cr
                            \null & \null \cr
                          {\ds k \over \ds k+q} & if $\atp > \ap$.\cr}
\label{eq:res}
\end{equation}
Here, $\atp$ and $\ap$ are the semimajor axes of the test particle and
the planet respectively.  From classical perturbation theory, the
strength of the resonance is roughly indicated by the order of the
commensurability $q$. On the uppermost panel of
Figure~\ref{fig:figone}, each resonance is labelled according to the
planet and the number $k$. For example, the resonance ME2 at semimajor
axis $\atp =0.142$ au with $q =7$ corresponds to a mean motion ration
of $9/2$ with Mercury. As seen from the Figure, some resonances tend
to excite the test particle's inclination (ME1, $q=6$), while others
pump up the eccentricity (ME1, $q=4$; ME2, $q=7$; ME1, $q=3$; ME3,
$q=7$). The effect of the Mercurian resonances occurs over a
relatively broad span of semimajor axes, (especially near ME1, $q=3$)
probably because of the wide range of eccentricity ($0.1 < e< 0.3$)
and inclination ($ 0^\circ < i < 10.5^\circ$) that Mercury undergoes
over the 100 Myr integration timespan. Conversely, the effects of the
Venusian resonances (V1, $q=7$; V1, $q=6$) appear much more localized
at semimajor axes 0.181 au and 0.198 au and excite both the
eccentricity and the inclination of the test particle. 

The collisional time scale in the Vulcanoid belt is short because the
volume of space is small and because the relative velocities are high.
Any sizeable population of bodies must have evolved through mutual
collisions.  Numerical models of the collisional processes (Stern \&
Durda 2000) suggest that a population of at most a few hundred objects
with radii larger than 1 km could have survived from primordial times.
As collisional evolution proceeds most quickly at smaller heliocentric
distances, Stern \& Durda suggest that the most favourable location to
search is nearly circular orbits near the very outer edge of the
dynamical stable Vulcanoid belt ($\sim 0.2$ au). However,
Figure~\ref{fig:figone} shows that the Venusian resonance (V1, $q=6$)
at 0.198 au excites both the eccentricity and inclination of such
objects. Rather, it is the region slightly inward between $0.16$ and
$0.18$ au that is the most likely to contain surviving objects. In
other words, the optimum location in which to search for Vulcanoids is
$9^\circ \lta |\ell_{\rm g} | \lta 10^\circ$ and $| \beta_{\rm g} |
\lta 1^\circ$, where ($\ell_{\rm g},\beta_{\rm g}$) are geocentric
ecliptic longitude and latitude.

Very recently, two searches using the Large Angle Spectroscopic
Coronagraph (LASCO) on the {\it SOHO} satellite have been conducted
(Durda et al. 2000; Schumacher \& Gay 2001). In particular, Durda et
al. found no Vulcanoids to a moving object detection limiting
magnitude of $V \sim 8$, corresponding to objects $\sim 50$ km in
diameter.  There have also been a number of daytime searches with
infrared telescopes (Leake et al. 1987; Campins et al. 1996), although
limited to small fields of view.  The closeness of Vulcanoids to the
Sun causes them to be hot. The search strategy exploits the fact that
Vulcanoids have a substantial component of infrared emission with the
K and L bands offerring the best chances of detection. Leake et al.
reached a limiting magnitude of $L \sim 5$ but covered only $\sim 6$
square degrees of the sky.  The new infrared wide-field telescope
VISTA~\footnote{http:www.vista.ac.uk} has a field of view of 0.5
square degrees in the K band and will be ideal for deep Vulcanoid
searches in the ecliptic centered around $9^\circ \lta |\ell_{\rm g} |
\lta 10^\circ$.

The European Space Agency (ESA) satellite {\it
BepiColombo}~\footnote{http://sci.esa.int/home/bepicolombo/} is
scheduled for lauch in 2009. Its main objective is to orbit and image
the planet Mercury. However, it will also be equipped with a small
telescope (20-30cm) and CCD camera for asteroid detection. The
telescope points in the direction of motion of the satellite, which is
in a polar orbit around the planet.  Since the satellite is
continuously looking at the surface of Mercury, the telescope is
rotating with respect to the stars. The field of view of the telescope
scans a great circle on the sky every $\sim 2.4$ hours and all objects
brighter than $V \sim 18$ crossing the scan region may be
detected. The telescope will image the area interior to Mercury but it
gets no closer to the Sun than $\sim 0.25$ au. As
Figure~\ref{fig:figone} shows, many of the objects in the outermost
part of the belt are on eccentric orbits with aphelion $Q > 0.25$ au
and so will spend some time in the scan region.  Therefore, {\it
BepiColombo} may detect some objects belonging to the outer part of
the Vulcanoid belt, although it will not image the bulk of the belt.

%
\vspace{1cm} 
\begin{table*}
\begin{tabular}{|c|c|c|c|c|c|c|}
\hline
Name &q &$Q$ &$i$ &$e$ &$a$ & Discoverer \\ 
\hline\hline
1989 ML & 1.099 & 1.447 & $4.4^\circ$ & 0.137 & 1.273 & Helin \& Alu \\
1993  HA & 1.094 & 1.463 & $7.7^\circ$ & 0.144 & 1.278 & Spacewatch \\
1996 XB27 & 1.120 & 1.258 & $2.5^\circ$ & 0.058 & 1.189 & Spacewatch \\
1998 HG49 & 1.065 & 1.335 & $4.42^\circ$ & 0.113 & 1.200 & Spacewatch \\
1998 KG3 & 1.024 & 1.298 & $5.5^\circ$ & 0.118 & 1.161 & Spacewatch \\
2000 AE205 & 1.004 & 1.322 & $4.5^\circ$ & 0.137 & 1.163 & LINEAR \\
2001 KW18 & 1.046 & 1.439 & $7.2^\circ$ & 0.158 & 1.243 & LONEOS \\
2001 SW169 & 1.184 & 1.313 & $3.6^\circ$ & 0.052 & 1.248 & LINEAR \\
1993 KA & 1.008 & 1.502 & $6.0^\circ$ & 0.197 & 1.255 & LINEAR \\
\hline
1998 YB & 1.222 & 1.420 & $6.8^\circ$ & 0.075 & 1.321 & LONEOS \\
2001 AE2 & 1.239 & 1.460 & $1.7^\circ$ & 0.082 & 1.350 & LINEAR \\
2001 QE96 & 1.274 & 1.347 & $7.3^\circ$ & 0.028 & 1.311& Spacewatch \\
\hline
\hline
\end{tabular}
\caption{Data on all asteroids between the semimajor axes of the Earth
and Mars with inclinations $i < 10^\circ$ and eccentricities $e < 0.2$
that are not planet-crossing. The object name, perihelion $q$,
aphelion $Q$, inclination $i$, eccentricity $e$ and semimajor axis $a$
are given, as well as the discovery team.  The upper nine asteroids
lie in the Earth-Mars belt, the lower three do not. (The source of the
data is The Minor Planet Center).}
\label{table:newdata}
\end{table*}

\section{The Earth-Mars Belt}

Let us now turn to the Earth-Mars belt. The high resolution simulation
between 1.08 and 1.28 au exhibits a rich palette of dynamical
behaviour. Figure~\ref{fig:figtwo} shows how the Earth-Mars belt has
been sculpted by various mean motion and secular resonances. In the
two lower panels, the filled circles show the instantaneous values at
100 Myr, while the crosses represent the maxima.  For mean motion
resonances, the upper panel indicates the semimajor axis of the
resonance and order of the commensurability $q$. For the secular
resonances, the upper panel shows the semimajor axis and inclination
at which the effects of the two secular frequencies $g_6$ and $s_6$
are greatest. These linear secular resonances have been extracted from
Michel \& Froeschl\'e (1997) and Williams \& Faulkner (1981). At the
$g_6$ (or $s_6$) secular resonances, the averaged precession frequency
of the asteroid's longitude of pericentre (or longitude of node)
becomes equal to the sixth eigenfrequency of the planetary system,
which is primarily due to Saturn.

The part of the belt between 1.08-1.117 au is populated with extremely
stable test particles. This can be seen from the mean values of the
instantaneous and maximum eccentricities ($\langle \einst \rangle =
0.037$, $\langle \emax \rangle = 0.082$) and inclinations ($\langle
\iinst \rangle = 1.81^\circ$ and $\imax = 4.86^\circ$). The gaps and
the spikes in eccentricity and inclination are in good agreement with
the location of the mean motion resonances.  However, from 1.17 au
outwards, the presence of the linear secular resonances with Saturn
tends to destabilize the orbits of the low inclination test
particles. An island of stability between 1.208 to 1.214 au is able to
avoid the disturbing effect of the secular resonances, thanks to the
fact that most of the test particles' inclinations lie in the safety
interval $1.8^\circ -6.4^\circ$. Objects between 1.214 au and 1.28 au
do not survive in this high resolution simulation. Nonetheless, Evans
\& Tabachnik (1999) found examples of some surviving test particles
out to 1.28 au in their original 100 Myr integrations which covered
the whole of the inner Solar system. We therefore take 1.28 au as the
outermost edge of the Earth-Mars belt.

Conventionally speaking, the members of the Earth-Mars belt fall under
the classification of Amor asteroids ($1.017 <q< 1.3$), which are part
of the population of Near-Earth Objects (NEOs). Ground-based programs
have found $\sim 2000$ NEOs, of which $\sim 800$ are Amors.  The total
number of NEOs is unknown. The present population is believed to have
arisen from at least three sources, namely asteroids ejected from the
Main Belt by the 3:1 mean motion resonance with Jupiter and the $s_6$
secular resonance; asteroids on Mars crossing orbits adjacent to the
Main Belt; and defunct comets that have evolved from the Kuiper Belt
through planetary encounters (e.g., Bottke et al. 2000).  Ejected Main
Belt asteroids and evolved comets are always planet-crossing and hence
show large fluctuations in their eccentricity. For example, Gladman et
al. (1997) show a number of dynamical tracks of asteroids ejected from
the Main Belt and their evolution is almost always towards greater
eccentricity. In one or two cases studied by Gladman et al., the
evolution does pass through low eccentricity phases, but usually only
for comparatively short periods of time. The probability of observing
this part of the evolution is low. So, NEOs on nearly circular orbits
are excellent candidates for being primordial.  In other words, if the
Earth-Mars belt of remnant planetesimals exists, an enhancement of
asteroids on nearly circular orbits is expected in this region. Evans
\& Tabachnik (1999) carried out a search through Bowell's (1989)
asteroidal orbital element database for objects between the semimajor
axes of Earth and Mars which are not planet-crossing and with low
inclinations and eccentricities. They found only three objects (1996
XB27, 1998 HG49 and 1998 KG3) among all the asteroids in the database.
All three lie between 1.08 and 1.28 au. The success of ground-based
surveys of NEOs in recent years has led to a substantial increase in
available data, so it is worthwhile repeating the calculations. Using
data from The Minor Planet Center, we searched for all objects between
the semimajor axes of Earth and Mars which are not planet-crossing and
which satisfy $i <10^\circ$ and $e < 0.2$.  The results are recorded
in Table~\ref{table:newdata}. There are twelve such objects known, of
which nine lie in the suggested Earth-Mars belt. This tends to support
the conjecture of Evans \& Tabachnik (1999) that there is an
enhancement of asteroids on nearly circular orbits in the belt.

Nonetheless, complex selection effects are present in the NEO
dataset. To decide the issue, a catalogue of NEOs with a
well-understood selection function is needed.  Fortunately, the ESA
scanning satellite {\it GAIA}
\footnote{http://astro.estec.esa.nl/gaia} will provide exactly this.
{\it GAIA} has three telescopes, ASTRO-1 and ASTRO-2 which perform
astrometry and broad-band photometry, and SPECTRO which performs
spectroscopy and medium-band photometry.  {\it GAIA} can detect all
objects brighter than $V \approx 20$.  For the NEO population, Mignard
(2002) shows that objects with absolute magnitude $H \lta 18$ will be
observed at least two times in the ASTRO or the SPECTRO fields of view
with the speed well-determined from the crossings. So, {\it GAIA} will
provide a reasonably complete inventory of the NEO population with
diameters of the order of a kilometre or larger.

\section{Conclusions}

Detailed maps of the fine-structure of two long-lived belts of stable
orbits in the Solar system have been provided. This extends the
results presented in Evans \& Tabachnik (1999), which reported
evidence that belts of remnant planetesimals could survive for the age
of the Solar system.  The Vulcanoid belt lies interior to the orbit of
Mercury and extends from 0.09 to 0.20 astronomical units (au), though
with a gaps at 0.15 and 0.18 au corresponding to a de-stabilising mean
motion resonances with Mercury and Venus.  Collisions drive evolution
in the Vulcanoid belt and this proceeds slower at larger heliocentric
distances.  Searches should concentrate on the region between
$0.16-0.18$ au where kilometre-sized Vulcanoids are most likely to be
found. The Earth-Mars belt between 1.08-1.17 au is an extremely stable
zone.  It is interrupted at 1.212 au due to the 3:4 commensurability
with the Earth, while secular resonances with Saturn can pump up the
eccentricities beyond 1.17 au. The belt may extend out to 1.28 au.

There are examples of stable orbits with small eccentricities and
inclinations which are nonetheless depleted of bodies (in the outer
Main Belt, for instance).  The populations of small bodies of the
solar system may have been dynamically excited or ejected by unknown
processes during the primordial evolution.  So the belts are worth
searching for because their presence or absence bears strongly on why
the inner Solar system looks the way it does today. For example,
substantial migration of Mercury would be strongly constrained by the
discovery of a population of asteroids in the Vulcanoid belt.

\section*{Acknowledgments}
We thank Stephano Mottola for information on {\it BepiColombo}. NWE is
supported by the Royal Society.

{}

\label{lastpage}
\end{document}